\documentclass{article}
\usepackage{spconf,amsmath,graphicx,amsfonts,epsfig,epstopdf}
\usepackage{times}
\usepackage{amsbsy}
\usepackage{latexsym}
\usepackage{amssymb}
\ninept
\begin{document}

\title{Study of Joint MMSE Consensus and Relay Selection Algorithms \\ for Distributed Beamforming}

\name{Hang Ruan * ~and Rodrigo C. de Lamare*$^\#$ \vspace{-0.5em} }

\address{ *Department of Electronics, The University of York, England, YO10 5BB\\
{ $^\#$CETUC, Pontifical Catholic University of Rio de Janeiro, Brazil}\\
 Emails: hr648@york.ac.uk,  delamare@cetuc.puc-rio.br \vspace{-0.5em}
 \sthanks{This work was supported in part by The University of York}}

\maketitle

\begin{abstract}
This work presents joint minimum mean-square error (MMSE) consensus algorithm
and relay selection algorithms for distributed beamforming. We propose joint
MMSE consensus relay and selection schemes with a total power constraint and
local communications among the relays for a network with cooperating sensors.
We also devise greedy relay selection algorithms based on the MMSE consensus
approach that optimize the network performance. Simulation results show that
the proposed scheme and algorithms outperform existing techniques for
distributed beamforming.
\end{abstract}


\section{Introduction}

Distributed beamforming has been widely investigated in wireless communications
in recent years \cite{r1,r2,r3,r15}. It is key for situations in which the
channels between the sources and the destination have poor quality so that
devices cannot communicate directly and the destination relies on relays that
receive and forward the signals \cite{r2}. The work in \cite{r3} formulates an
optimization problem that maximizes the output
signal-to-interference-plus-noise ratio (SINR) under the individual relay power
constraints. The approach in \cite{r7} proposes an MMSE consensus cooperative
relay networking scheme to exchange data among all the relays under a total
power constraint, which limits the total power of all relays regardless of the
power allocation. While local communications among the relays are enabled, the
ability to mitigate fading effects in wireless channels of the network can be
improved \cite{r8}. Further earlier works in \cite{r12} and \cite{r13} explored
local communications, while avoiding network centralized processing, which is
not desirable and always comes along with the use of total power constraints
\cite{r7}.

However, in most scenarios relays are either not ideally distributed
in terms of locations or the channels involved with some of the
relays have poor quality. Possible solutions can be categorized in
two approaches. One is to adaptively adjust the power of each relay
according to the qualities of its associated channels, known as
adaptive power control or power allocation. Some power control
methods based on channel magnitude and relative analysis has been
studied in \cite{r6,r14}. An alternative solution is to use relay
selection, which selects a number of relays according to a criterion
of interest while discarding the remaining relays. In \cite{r4},
multi-relay selections algorithm have been developed to maximize the
secondary receiver in a two-hop cognitive relay network. In
\cite{r8}, several optimum single-relay selection schemes and a
multi-relay selection scheme using relay ordering based on
maximizing the output SNR under individual relay power constraints
are developed and discussed. The work in \cite{r9} proposed a
low-cost greedy search method for the uplink of cooperative direct
sequence code-division multiple access systems, which approaches the
performance of an exhaustive search. Other approaches include the
use of subspace techniques
\cite{scharf,bar-ness,pados99,reed98,hua,goldstein,santos,qian,delamarespl07,xutsa,delamaretsp,kwak,xu&liu,delamareccm,wcccm,delamareelb,jidf,delamarecl,delamaresp,delamaretvt,jioel,rrdoa,delamarespl07,delamare_ccmmswf,jidf_echo,delamaretvt10,delamaretvt2011ST,delamare10,fa10,lei09,ccmavf,lei10,jio_ccm,ccmavf,stap_jio,zhaocheng,zhaocheng2,arh_eusipco,arh_taes,dfjio,rdrab,locsme,okspme,dcg_conf,dcg,dce,drr_conf,dta_conf1,dta_conf2,dta_ls,damdc,song,wljio,barc,jiomber,saalt}.
and large sensor arrays
\cite{mmimo,wence,Costa,delamare_ieeproc,TDS_clarke,TDS_2,switch_int,switch_mc,smce,TongW,jpais_iet,TARMO,badstc,baplnc,keke1,kekecl,keke2,wl_prec,Tomlinson,dopeg_cl,peg_bf_iswcs,gqcpeg,peg_bf_cl,Harashima,mbthpc,zuthp,rmbthp,Hochwald,BDVP},\cite{delamare_mber,rontogiannis,delamare_itic,stspadf,choi,stbcccm,FL11,jio_mimo,peng_twc,spa,spa2,jio_mimo,P.Li,jingjing,did,bfidd,mbdf}.

In this work, we propose a joint minimum mean-square error (MMSE) consensus and
relay selection approach and develop iterative greedy search based relay
selection algorithms for distributed beamforming. In the first proposed
algorithm we aim to find largest minimum value of a cost function regarding the
desired signal by enabling only one relay at once, testing through all the
relays one by one, then we select all the disabled relays when the network has
the largest MMSE (LMMSE) value in each iteration, namely, the SMMSE consensus
greedy (LMMSEC-G) relay selection algorithm. In the second proposed approach,
we aim to find the smallest minimum value of the same cost function, by
disabling only one relay at once, testing through all the relays one by one,
then preserve all the selected relays when the network has the smallest MMSE
(SMMSE) estimate in each iteration, namely, the the SMMSE consensus greedy
(SMMSEC-G) relay selection algorithm. We compare the proposed greedy relay
selection algorithms to the exhaustive search and the scenario without relay
selection, showing their excellent output SINR performance which is close to
the exhaustive search approach.

This paper is organized as follows. In Section 2, the system model is
introduced. In Section 3, the joint MMSE consensus and relay selection scheme
is presented. In Section 4, the greedy relay selection algorithms are proposed.
In Section 5, simulations are presented and discussed. Finally, conclusions are
drawn in Section 6.

\section{System Model}

\begin{figure}[!htb]
\begin{center}
\def\epsfsize#1#2{0.85\columnwidth}
\epsfbox{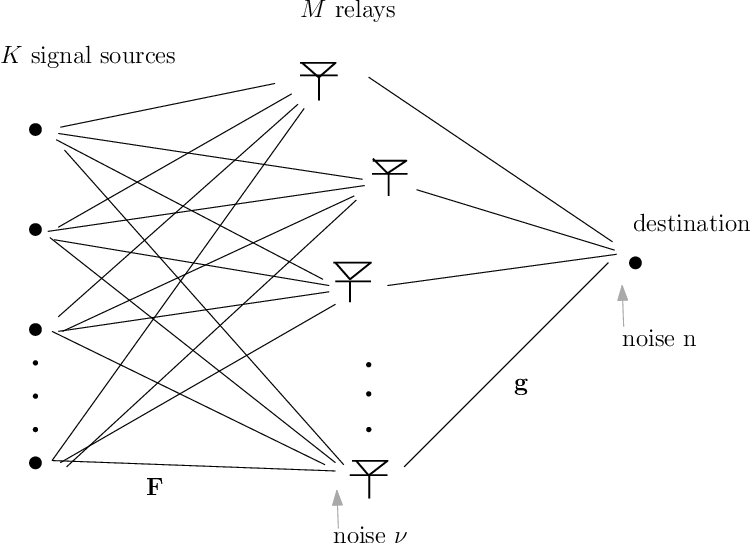} \vspace{-1em} \caption{System model.}
\label{model}
\end{center}
\end{figure}

We consider a wireless communication network consisting of $K$ signal sources
(one desired signal with the others as interferers), $M$ distributed
single-antenna relays and a destination. It is assumed that the quality of the
channels between the signal sources and the destination is very poor so that
direct communications is not possible and their links are negligible. The $M$
relays receive information transmitted by the signal sources and then
retransmit to the destination as a beamforming procedure, in which a two-step
amplify-and-forward (AF) protocol (as shown in Fig. \ref{model}) is considered
as required for cooperative communications.

In the first step, the sources transmit the signals to the relays as
\begin{equation}
{\bf x}={\bf F}{\bf s}+{\boldsymbol \nu}, \label{eq1}
\end{equation}
where ${\bf s}=[s_1, s_2, \dotsb, s_K] \in {\mathbb C}^{1 \times K}$ are signal
sources with zero mean, $[.]^T$ denotes the transpose, $s_k=\sqrt{P_k}s$,
$E[|s|^2]=1$, $P_k$ is the transmit power of the $k$th signal source,
$k=1,2,\dotsb,K$, $s$ is the information symbol. Without loss of generality we
can assume $s_1$ as the desired signal while the others are treated as
interferers. ${\bf F}=[{\bf f}_1, {\bf f}_2, \dotsb, {\bf f}_K] \in {\mathbb
C}^{M \times K}$ is the channel matrix between the signal sources and the
relays, ${\bf f}_k=[f_{1,k}, f_{2,k}, \dotsb, f_{M,k}]^T \in {\mathbb C}^{M
\times 1}$, $f_{m,k}$ denotes the channel between the $m$th relay and the $k$th
source ($m=1,2, \dotsb, M$, $k=1,2,\dotsb, K$). ${\boldsymbol \nu}=[\nu_1,
\nu_2, \dotsb, \nu_M]^T \in {\mathbb C}^{M \times 1}$ is the complex Gaussian
noise vector at the relays and $\sigma_{\nu}^2$ is the noise variance at each
relay ($\nu_m$ \~{} ${\mathcal CN}(0,\sigma_{\nu}^2)$). The vector ${\bf x} \in
{\mathbb C}^{M \times 1}$ represents the received data at the relays. In the
second step, the relays transmit ${\bf y} \in {\mathbb C}^{M \times 1}$ which
is an amplified and phase-steered version of ${\bf x}$, which can be written as
\begin{equation}
{\bf y}={\bf W}{\bf x}, \label{eq2}
\end{equation}
where ${\bf W}={\rm diag}[w_1, w_2,  \dotsb, w_M] \in {\mathbb C}^{M \times M}$
is a diagonal matrix whose diagonal entries denote the beamforming weights. The
signal received at the destination is given by
\begin{equation}
z={\bf g}^T{\bf y}+n, \label{eq3}
\end{equation}
where $z$ is a scalar, ${\bf g}=[g_1, g_2, \dotsb, g_M]^T \in {\mathbb C}^{M
\times 1}$ is the complex Gaussian channel vector between the relays and the
destination, $n$ ($n$ \~{} ${\mathcal CN}(0,\sigma_n^2)$,
$\sigma_n^2=\sigma_{\nu}^2$) is the noise at the destination and $z$ is the
received signal at the destination.

Note that both ${\bf F}$ and ${\bf g}$ are modeled as Rayleigh distributed
(i.e., both the real and imaginary coefficients of the channel parameters have
Gaussian distribution). Using the Rayleigh distribution for the channels, we
also consider distance based large-scale channel propagation effects that
include distance-based fading (or path loss) and shadowing. Distance-based
fading represents how a signal is attenuated as a function of the distance and
can be highly affected by the environment. \cite{r10,r11} An exponential based
path loss model can be described by
\begin{equation}
\gamma=\frac{\sqrt{L}}{\sqrt{d^{\rho}}}, \label{eq4}
\end{equation}
where $\gamma$ is the distance based path loss, $L$ is the known path loss at
the destination, $d$ is the distance of interest relative to the destination
and $\rho$ is the path loss exponent, which can vary due to different
environments and is typically set within $2$ to $5$, with a lower value
representing a clear and uncluttered environment which has a slow attenuation
and a higher value describing a cluttered and highly attenuating environment.
Shadow fading describes the phenomenon where objects can obstruct the
propagation of the signal attenuating the signal further, and can be modeled as
a random variable with probability distribution given by
\begin{equation}
\beta=10^{(\frac{\sigma_s{\mathcal N}(0,1)}{10})}, \label{eq5}
\end{equation}
where $\beta$ is the shadowing parameter, ${\mathcal N}(0,1)$ means the
Gaussian distribution with zero mean and unit variance, $\sigma_s$ is the
shadowing spread in dB. The shadowing spread reflects the severity of the
attenuation caused by shadowing, and is typically given between $0$dB to $9$dB.
The channels modeled with both path-loss and shadowing are described by
\begin{equation}
{\bf F}=\gamma\beta{\bf F}_0,  \label{eq6}
\end{equation}
\begin{equation}
{\bf g}=\gamma\beta{\bf g}_0,  \label{eq7}
\end{equation}
where ${\bf F}_0$ and ${\bf g}_0$ denote the Rayleigh distributed channels
without path-loss and shadowing \cite{r10,r11}.

\section{Proposed Joint MMSE Consensus and Relay Selection}

In this section, we detail the proposed joint MMSE consensus and relay
selection scheme for distributed beamforming using an alternating optimization
approach in which the relay selection is followed by MMSE consensus of
beamformers. We assume at the $m$th relay the MMSE estimate of the desired
signal ${\hat s}_{m,1}$ can be found as
\begin{equation}
{\hat s}_{m,1}={\phi}_mx_m, \label{eq8}
\end{equation}
where $${\phi}_m={\rm
arg~min}_{\phi}E[|s_1-{\phi}x_m|^2]=\frac{f_{m,1}^*P_1}{\sum_{k=1}^K|f_{m,k}|^2P_k+\sigma_n^2}.$$
For convenience we define ${\tilde s}_{m,1}=\frac{{\hat s}_{m,1}}{E[|{\hat
s}_{m,1}|^2]}$ and the normalized relay weight as $\frac{w_m\phi_m}{E[|{\hat
s}_{m,1}|^2]}$, so that the total transmission power can be expressed as
$\sum_{m=1}^ME[|w_m{\tilde s}_{m,1}|^2]=\sum_{m=1}^M|w_m|^2$. Therefore, the
MMSE consensus optimization associated with a fixed set of relays under a total
power constraint is given by
\begin{equation}
\begin{aligned}
w_m^* = \arg \min_{w_m}~~ \sum_{m=1}^ME[|s_1-g_mw_m{\tilde s}_{m,1}|^2] \\
s.t. ~ \sum_{m=1}^M|w_m|^2 \leq P_T, \label{eq9}
\end{aligned}
\end{equation}
where $P_T$ is the maximum allowable total transmit power of all relays. The
relay selection problem for the MMSE consensus can be described as an
optimization problem using a total relay transmit power constraint described by
\begin{equation}
\begin{aligned}
{\mathcal S}_{\rm opt}=\underset{\boldsymbol\alpha, {\bf w}}{\rm arg~min}~~MMSE({\mathcal S}) \\
s.t. \sum_{m=1}^M\alpha_m|w_m|^2 \leq P_T, \\
\alpha_m \in \{0,1\},m=1,2,\dotsb,M  \label{eq10}
\end{aligned}
\end{equation}
where $MMSE({\mathcal S})=\sum_{m= 1}^M\alpha_mE[|s_1-g_mw_m{\tilde
s}_{m,1}|^2]$, ${\bf w}=[w_1, w_2,  \dotsb, w_M]^T \in {\mathbb C}^{M \times
1}$, ${\mathcal S}_{\rm opt}$ and ${\mathcal S}$ are the optimum relay set of
size $M_{\rm opt}$, ($1\leq{M_{\rm opt}}\leq{M}$) and the original relay set of
size $M$, respectively. The vector ${\boldsymbol \alpha}=[\alpha_1, \alpha_2,
\dotsb,\alpha_M]^T \in {\mathbb R}^{M \times 1}$, $\alpha_m$ ($m=1,\dotsb,M$)
is the relay cooperation parameter vector which determines if the $m$th relay
will cooperate. The solution of \eqref{eq10} regarding $w_m$ indicates the
following relationship:
\begin{equation}
w_m^*=\frac{\alpha_mg_m^*}{\lambda
+\alpha_m|g_m|^2}\sqrt{\frac{|f_{m,1}|^2P_1^2}{\sum_{k=1}^K|f_{m,k}|^2P_k+\sigma_n^2}},
\label{eq11} \\
\end{equation}
where $\lambda$ is the Langrange multiplier, which can be determined by
enabling the local communication of the relays with an MMSE consensus approach.
To this end, we employ an auxiliary beamforming weight vector $\tilde{\bf
w}_m=[{\tilde w}_{1,m}, \dotsb, {\tilde w}_{M,m}]^T$ and consider the following
joint optimization problem:
\begin{equation}
\begin{aligned}
\min_{\tilde{\bf w}_m,\boldsymbol\alpha}~~ \alpha_mE[|s_1-g_mw_m{\tilde s}_{m,1}|^2] \\
s.t. ~ ||\boldsymbol\alpha||_1||\tilde{\bf w}_m||^2 \leq P_T, \tilde{\bf w}_m={\bf w}, \\
\alpha_m \in \{0,1\},m=1,2,\dotsb,M   \label{eq12}
\end{aligned}
\end{equation}
where ${\bf w}={\rm diag}({\bf W}) \in {\mathbb C}^{M \times 1}$. It is
supposed that the $m$th relay is connected to a subset of relays denoted by
${\mathcal M}_m$. The second constraint in \eqref{eq12} can be replaced by
$\tilde{\bf w}_m=\tilde{\bf w}_q, q\in{\mathcal M}_m$ so that \eqref{eq12} is
reformulated as
\begin{equation}
\begin{aligned}
{\rm min}\underset{\tilde{\bf w}_m,\boldsymbol\alpha}~~\alpha_mE[|s_1-g_mw_m{\tilde s}_{m,1}|^2]+\lambda_m(i)(||\boldsymbol\alpha||_1||\tilde{\bf w}_m||^2-P_T) \\
+\sum_{q\in{\mathcal M}_m}{\bf\tau}_{m,q}^T(\tilde{\bf w}_m-\tilde{\bf w}_q),
\label{eq13}
\end{aligned}
\end{equation}
where $\lambda_m(i)$ and ${\bf\tau}_{m,q}$ are Lagrange multipliers.The proposed algorithmic solution relies on the alternating optimization associated with relay selection, computation of the optimal weights and Lagrange
multipliers at the $m$th relay as
\begin{equation}
{\tilde w}_{t,m}=
\begin{cases}
\frac{\alpha_mg_m^*}{\lambda_m(i)
+\alpha_m|g_m|^2}(\sqrt{\frac{|f_{m,1}|^2P_1^2}{\sum_{k=1}^K|f_{m,k}|^2P_k+\sigma_n^2}}-\frac{\sum_{q\in{\mathcal M}_m}\tau_{m,q;m}}{2}), \\
\text{if} ~~ t=m \\
-\frac{\alpha_m\sum_{q\in{\mathcal M}_m}\tau_{m,q;t}}{2\lambda_m(i)}, \\ \text{if} ~~ t
\neq m \label{eq14}
\end{cases}
\end{equation}
where $\tau_{m,q;t}$ denotes the $t$th element of ${\bf\tau}_{m,q}$. The
Lagrange multipliers are updated as follows
\begin{equation}
\lambda_m(i)=|\lambda_m(i-1)+\mu_{\lambda}(||\tilde{\bf w}_m||^2-P_T)|, \label{eq15}
\end{equation}
\begin{equation}
{\bf\tau}_{m,q}(i)={\bf\tau}_{m,q}(i-1)+\mu_{\tau}({\bf u}_m-{\bf u}_q), \label{eq16}
\end{equation}
where $\mu_{\lambda}$ and $\mu_{\tau}$ are step sizes with small positive
values, ${\bf u}_m=[|w_{1,m}|, \dotsb, |w_{M,m}|]^T$ and $i$ is the time index \cite{r7}.

\section{Proposed Greedy Relay Selection Algorithms}

In this section, we detail the algorithms that perform relay selection, develop the LMMSEC-G and SMMSEC-G relay selection algorithms and review the exhaustive search.

\subsection{{LMMSEC-G relay selection algorithm}}

We develop the LMMSEC-G algorithm to obtain the solution of \eqref{eq12}, which
depends on the relay selection parameter vector $\boldsymbol\alpha$ to find the
optimum $\boldsymbol\alpha$. The LMMSEC-G works in an iterative way and
discards only the worst relay to find the optimal relay set in each iteration.
Additionally, the parameter $M_{min}$ can be introduced to restrict the minimum
number of relays that must be used. The LMMSEC-G algorithm finds the largest
MMSE at each iteration and selects the complementary relays, which is described
as follows
\begin{equation}
\begin{aligned}
\bar{\mathcal S}(i)=\underset{\boldsymbol\alpha(i)}{\rm arg~max} ~~ MMSE({\mathcal S}(i-1)) \\
s.t. ~ ||\tilde{\bf w}_m(i)||^2 = \leq P_T, \tilde{\bf w}_m(i)={\bf w}(i), \\
\alpha_m(i) \in \{0,1\},m=1,2,\dotsb,M \\
M-i\geq{M_{min}},  \label{eq17}
\end{aligned}
\end{equation}
where $\bar{\mathcal S}(i)$ denotes the complement of the set ${\mathcal S}(i)$
from set ${\mathcal S}(i-1)$. The optimization problem compares all the MMSE
values assuming that only one different single relay is enabled while the
others are disabled. LMMSEC-G cancels the relay with largest MMSE value from
set ${\mathcal S}(i-1)$ and evaluates the MMSE performance of the remaining
relays, which is solved only once in each iteration. If the MMSE in the current
iteration is smaller than that in the previous iteration (i.e.
$MMSE(i)<MMSE(i-1)$), then the selection process continues; if $MMSE(i) \geq
MMSE(i-1)$, we cancel the selection of the current iteration and keep the relay
set ${\mathcal S}(i-1)$ and $MMSE(i-1)$. The LMMSEC-G algorithm is shown in
Table. \ref{table1}.

{\footnotesize
\begin{table}
\begin{center}
\caption{LMMSEC-G relay selection algorithm}
\begin{tabular}{l}
\hline
Initialize ${\mathcal S}_{opt}={\mathcal S}(0)$, $\boldsymbol\alpha={\bf 1}$, $\tilde{\bf w}_m={\bf 1}$, \\
$\lambda_m(0)=1$, $\tau_{m,q}(0)=1$ and compute $MMSE_{o}=MMSE(0)$. \\
for $i=1,\dotsb,M-M_{min}$ \\
{\bf step~1}: \\
compute ${\bf\tau}_{m,q}(i)={\bf\tau}_{m,q}(i-1)+\mu_{\tau}({\bf u}_m-{\bf u}_q)$, \\
and ${\bf\tau}_{m,q}(i)={\bf\tau}_{m,q}(i-1)+\mu_{\tau}({\bf u}_m-{\bf u}_q)$. \\
{\bf step~2}: \\
compute the consensus weight using \eqref{eq14}. \\
{\bf step~3}: \\
~~~~solve the optimization problem \eqref{eq17} and obtain $\bar{\mathcal S}(i)$. \\
~~~~compute ${\mathcal S}(i)$ using ${\mathcal S}(i-1)$-$\bar{\mathcal S}(i)$. \\
~~~~compute $MMSE(i)$ using ${\mathcal S}(i)$. \\
~~~~compare $MMSE(i)$ to $MMSE(i)(i-1)$, \\
~~~~if $MMSE(i)<MMSE(i-1)$ \\
~~~~~~~~update ${\mathcal S}_{opt}={\mathcal S}(i)$ and $MMSE_{o}=MMSE(i)$. \\
~~~~~~~~update $\boldsymbol\alpha(i)$. \\
~~~~else \\
~~~~~~~~keep ${\mathcal S}_{opt}={\mathcal S}(i-1)$ and $MMSE_{o}=MMSE(i-1)$. \\
~~~~~~~~break. \\
~~~~end if. \\
end for. \\
\hline
\end{tabular} \label{table1}
\end{center}
\end{table}}

\subsection{{SMMSEC-G relay selection algorithm}}

The proposed SMMSEC-G algorithm is an alternative way to find the solution of
\eqref{eq12} regarding $\boldsymbol\alpha$, which aims to find the smallest
MMSE from the remaining relays after disabling a single relay each time. It is
an improved greedy search based method which also works in iterations but with
higher complexity and much better performance. We also consider $M_{min}$ as a
restriction to the minimum number of relays that must be used. Before the first
iteration all relays are considered (i.e., ${\mathcal S}(0)={\mathcal S}$).
Consequently, we solve the following problem once for each iteration in order
to cancel the relay with worst performance from set ${\mathcal S}(i-1)$ and
evaluate the $MMSE(i)$ at time instant i:
\begin{equation}
\begin{aligned}
{\mathcal S}(i)=\underset{\boldsymbol\alpha(i)}{\rm arg~min} ~~ MMSE({\mathcal S}(i-1)) \\
s.t. ~ (M-i)||\tilde{\bf w}_m(i)||^2 \leq P_T, \tilde{\bf w}_m(i)={\bf w}(i), \\
\alpha_m(i) \in \{0,1\},m=1,2,\dotsb,M \\
M-i\geq{M_{min}}, \label{eq18}
\end{aligned}
\end{equation}
If the MMSE in the current iteration is lower than that in the previous
iteration (i.e. $MMSE(i)<MMSE(i-1)$), then the selection process continues; if
$MMSE(i)\geq MMSE(i-1)$, we cancel the selection of the current iteration and
keep the relay set ${\mathcal S}(i-1)$ and $MMSE(i-1)$. The SMMSEC-G algorithm
is shown in Table. \ref{table2}.

\begin{table}
\begin{center}
\caption{SMMSECG}
\begin{tabular}{l}
\hline
Initialize ${\mathcal S}_{opt}={\mathcal S}(0)$, $\boldsymbol\alpha={\bf 1}$, $\tilde{\bf w}_m={\bf 1}$, \\
$\lambda_m(0)=1$, $\tau_{m,q}(0)=1$ and compute $MMSE_{o}=MMSE(0)$. \\
for $i=1,\dotsb,M-M_{min}$ \\
{\bf step~1}: \\
compute ${\bf\tau}_{m,q}(i)={\bf\tau}_{m,q}(i-1)+\mu_{\tau}({\bf u}_m-{\bf u}_q)$, \\
and ${\bf\tau}_{m,q}(i)={\bf\tau}_{m,q}(i-1)+\mu_{\tau}({\bf u}_m-{\bf u}_q)$. \\
{\bf step~2}: \\
compute the consensus weight using \eqref{eq14}. \\
{\bf step~3}: \\
~~~~solve the optimization problem \eqref{eq18} and obtain ${\mathcal S}(i)$. \\
~~~~compute $MMSE(i)$ using ${\mathcal S}(i)$. \\
~~~~compare $MMSE(i)$ to $MMSE(i-1)$, \\
~~~~if $MMSE(i)<MMSE(i-1)$ \\
~~~~~~~~update ${\mathcal S}_{opt}={\mathcal S}(i)$ and $MMSE_{o}=MMSE(i)$. \\
~~~~~~~~update $\boldsymbol\alpha(i)$. \\
~~~~else \\
~~~~~~~~keep ${\mathcal S}_{opt}={\mathcal S}(i-1)$ and $MMSE_{o}=MMSE(i-1)$. \\
~~~~~~~~break. \\
~~~~end if. \\
end for. \\
\hline
\end{tabular} \label{table2}
\end{center}
\end{table}

\subsection{{Exhaustive Search}}

In an exhaustive search procedure, we test every possible combinations among
all the relays, which means the change of status if a relay is chosen or not
will contribute to a different possible combination. To obtain the global
optimum solution, we need to run the consensus algorithm once without
iterations. Also, we can predefine $M_{fix}$ as the required selected number of
relays as an additional requirement. However, the complexity can be extremely
high depending on the number of relays.

\section{Simulations}

\begin{figure}[!htb]
\begin{center}
\def\epsfsize#1#2{0.9\columnwidth}
\epsfbox{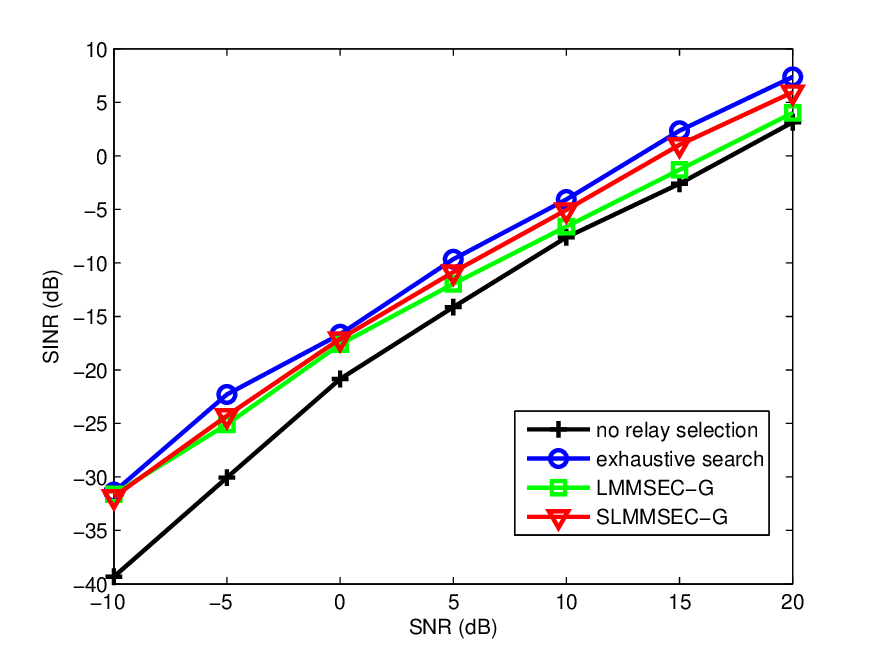}\vspace{-1.2em} \caption{SINR performance versus SNR}
\label{fig1}
\end{center}
\end{figure}

\begin{figure}[!htb]
\begin{center}
\def\epsfsize#1#2{0.9\columnwidth}
\epsfbox{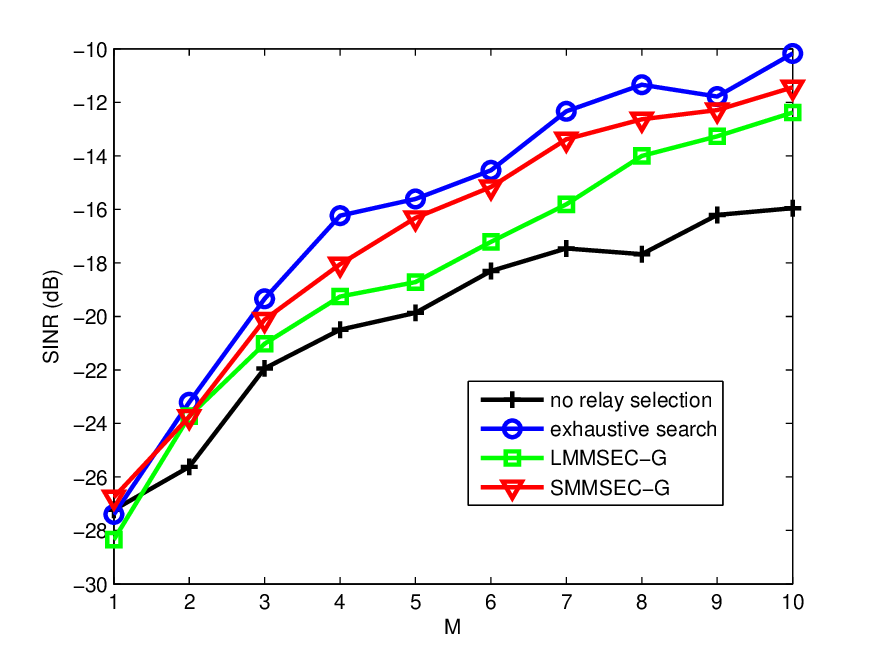}\vspace{-1.2em} \caption{SINR performance versus M}
\label{fig2}
\end{center}
\end{figure}

In the simulations we focus on the output SINR performance
comparison of the proposed LMMSEC-G and SMMSEC-G algorithms by
varying the input SNR or the total number of relays in the network.
The parameters used include: number of signal sources $K=3$, the
path loss exponent $\rho=2$, the power path loss from signals to the
destination $L=10$dB, shadowing spread $\sigma_s=3$dB, $P_T=1$dBW,
$M_{min}=1$. For the local communication between the relays, we set
${\mathcal M}_m=\{m+1\}$ and ${\mathcal M}={1}$,
$\mu_{\lambda}=\mu_{\tau}=0.001$. $100$ repetitions are executed for
each of the studied methods. In Fig. \ref{fig1}, we fixed the total
number of relays $M=5$ and interference-to-noise ratio (INR) at
$10$dB and evaluate the SINR versus SNR performance of the joint
MMSE consensus and relay selection approaches and the existing
techniques. Both the greedy search based methods, namely, LMMSEC-G
and SMMSEC-G, increase the SINR performance as compared with the
case without any relay selection and approach the exhaustive search
especially at low SNRs. Fig. \ref{fig2} illustrates that with a
fixed SNR($0$dB) and INR($0$dB) how the output SINR varies when the
total number of relays in the network increases. It is clear that
using more relays enhance the overall network performance and the
SMMSEC-G method performs very close to the exhaustive search. The
proposed techniques could also be evaluated in terms of BER
performance
\cite{dopeg_cl,peg_bf_iswcs,gqcpeg,peg_bf_cl,memd,delamare_mber,peng_twc,spa,jio_mimo,P.Li,jingjing,did,bfidd,mbdf,badstc2}.

\section{Conclusion}

We have proposed a joint MMSE consensus and relay selection approach and
developed efficient algorithms for distributed beamforming. We have proposed
the LMMSEC-G and SMMSEC-G greedy optimization algorithms based on the MMSE
criterion with known network quantities and relay selection strategies, which
determines if a relay should cooperate or not in the network. The LMMSEC-G and
SMMSEC-G algorithms have shown excellent performance and outperformed
previously reported techniques.

\end{document}